\documentclass{ws-mpla}
\def\citebk#1{\raisebox{-1.85mm}[0mm][0mm]{\Large\cite{#1}}}

\begin{document}

\markboth{V.P. Gusynin, V.A. Miransky and I.A. Shovkovy}
{Surprises in nonperturbative dynamics in $\sigma$-model at finite 
density}

%
\catchline{}{}{}{}{}
%

\title{SURPRISES IN NONPERTURBATIVE DYNAMICS IN $\sigma$-MODEL AT FINITE 
DENSITY}

\author{\footnotesize V.P. GUSYNIN}

\address{Bogolyubov Institute for Theoretical Physics,
03143, Kiev, Ukraine\\
vgusynin@bitp.kiev.ua}

\author{V.A. MIRANSKY\footnote{On leave from 
       Bogolyubov Institute for Theoretical Physics,
       03143, Kiev, Ukraine.}}

\address{Department of Applied Mathematics, University of Western
Ontario\\ 
London, Ontario N6A 5B7, Canada\\
vmiransk@uwo.ca}

\author{I.A. SHOVKOVY$^{*}$}

\address{Institut f\"{u}r Theoretische Physik,
       J.W. Goethe-Universit\"{a}t,
       D-60054 Frankurt/Main, Germany\\
shovkovy@th.physik.uni-frankfurt.de}

\maketitle

\pub{Received (Day Month Year)}{Revised (Day Month Year)}

\begin{abstract}

The linear $SU(2)_{L}\times SU(2)_R$ $\sigma$-model occupies a unique 
place in elementary particle physics and quantum field theory. It has 
been recently realized that when a chemical potential for hypercharge 
is added, it becomes a toy model for the description of the dynamics 
of the kaon condensate in high density QCD. We review recent results in
nonperturbative dynamics obtained in the ungauged and gauged versions of 
this model.

\end{abstract}

\ccode{PACS Nos.: 11.15.Ex, 11.30.Qc.}

\section{Introduction}

The linear $SU(2)_{L}\times SU(2)_R$ $\sigma$-model occupies 
a unique place in elementary particle physics and quantum field 
theory. Introduced back in 1960 by Gell-Mann and Levy,\cite{GML} 
the model is at the heart of particle and nuclear dynamics at 
very different scales. At the scale of order 10 MeV, the model with 
sigma, pion and nucleon fields is relevant for nuclear structure 
calculations. Then, at the scale of order 100 MeV, the model with 
nucleon fields replaced by quark ones, is an effective theory of 
chiral dynamics of hadrons (for a review, see Ref.~\citebk{HK}).
And at the scale of order 100 GeV, the model is nothing else but
the Higgs sector of the electroweak theory.    

In this brief review, we will describe new and quite surprising
phenomena taking place in the linear $SU(2)_{L}\times SU(2)_R$
$\sigma$-model at finite density or, more precisely, at the finite
chemical potential $\mu$ for the hypercharge $Y$. The interest to
this model is connected with that it is a toy
model for the description of the dynamics of the kaon 
condensate\cite{BS} 
in the color-flavor locked phase\cite{ARW} of high density QCD
that may exist in cores of compact stars.

We will consider dynamics at finite $\mu$
both in the linear $\sigma$-model itself\cite{MS,STV} 
and in its gauged 
version, when the subgroup $SU(2)_{L}\times U(1)_{Y}$ is being 
gauged.\cite{GMS} The central results are as follows:

\begin{enumerate}
   \item In the ungauged $\sigma$-model, the spontaneous breakdown of
$SU(2) \times U(1)_{Y}$ symmetry, caused by the chemical potential,
leads to a lesser number of
Nambu-Goldstone bosons than that required by the Goldstone
theorem. One of the consequences of this phenomenon is that
the system is not a superfluid
despite the presence of a condensate.

  \item In the gauged version of the model, the
spontaneous breakdown of
$SU(2) \times U(1)_{Y}$ symmetry, caused by the chemical potential,
is always accompanied by spontaneous breakdown of both rotational
symmetry and electromagnetic $U(1)_{em}$. 

  \item The spectrum of excitations
in the gauged model is very rich. In particular, there
exist excitation branches that behave as phonon-like quasiparticles
for small momenta and as roton-like ones for large momenta. These
roton-like excitations are present because of gauge vector fields.
This can shed light on microscopic nature of roton-like excitations, 
which is an old problem in the theory of 
superfluidity.\cite{L,F,superfluid} 
It also suggests that this model can be relevant for
superfluid and superconducting systems.
\end{enumerate}

\section{Dynamics with abnormal number of Nambu-Goldstone bosons}
\label{ng}

Recently a class of relativistic models with a finite density of 
matter has been revealed in which spontaneous
breakdown of continuous symmetries leads to a lesser number of
Nambu-Goldstone (NG) bosons than that required by the Goldstone
theorem.\cite{MS,STV} 
The simplest representative of this class is the linear
$SU(2)_{L}\times SU(2)_R$  $\sigma$-model with the chemical potential
for the hypercharge $Y$,
\begin{eqnarray}
{\cal L} = (\partial_{0} +i \mu )\Phi^{\dagger}
(\partial_{0} -i \mu )\Phi 
-\partial_{i}\Phi^{\dagger}
\partial_{i}\Phi -m^{2}\Phi^{\dagger} \Phi
-\lambda(\Phi^{\dagger} \Phi)^{2},
\label{L-model}
\end{eqnarray}
where $\Phi$ is a complex doublet field.
The 
chemical potential $\mu$ is provided by external conditions (to be specific, 
we take $\mu > 0$). For example, in the case of dense QCD with the kaon 
condensate, $\mu$ is
$\mu = m^{2}_s/2p_F$, where $m_s$ is the current mass of the 
strange quark and $p_F$ is the quark Fermi momentum.\cite{BS}
Note that the terms with the chemical potential
reduce the initial $SU(2)_{L}\times SU(2)_R$ symmetry   
to the $SU(2)_L \times U(1)_Y$ one. This follows from the fact that
the hypercharge generator $Y$ is $Y = 2I^{3}_{R}$ where $I^{3}_{R}$ 
is the
third component of the right handed isospin generator. Henceforth
we will omit the subscripts $L$ and $R$, allowing various
interpretations of the $SU(2)$ [for example, in the dynamics of
the kaon condensate, it is just the conventional
isospin symmetry $SU(2)_{I}$ and $\Phi^{T} = (K^{+},K^{0}$)].

The terms containing the chemical potential in Eq.~(\ref{L-model}) are
\begin{equation} 
i\mu \Phi^{\dagger}\partial_{0}\Phi - i\mu\partial_{0}
\Phi^{\dagger}\Phi + \mu^{2}\Phi^{\dagger} \Phi. 
\label{mu-terms}
\end{equation} 
The last term in this expression makes the mass term in Lagrangian
density (\ref{L-model}) to be $(\mu^{2} - m^2)\Phi^{\dagger} \Phi$. 
Therefore for supercritical values of the chemical potential,
$\mu^2 > m^2$, there is an instability resulting in the spontaneous
breakdown of $SU(2) \times U(1)_Y$ down to $U(1)_{em}$ connected with 
the electrical charge $Q_{em} =  I^{3} + \frac{1}{2}Y$. One may expect 
that this implies the existence of three NG bosons. The explicit
calculation shows that this is not the case. To see this, we represent 
the field in the following form:
\begin{equation}
\Phi=\left(\begin{array}{c} 0 \\ \varphi_{0}
\end{array}\right)
+\frac{1}{\sqrt{2}}\left(\begin{array}{c}
\varphi_{1}+i\varphi_{2} \\
\tilde{\varphi}_{1}+i\tilde{\varphi}_{2}
\end{array}\right),
\quad\mbox{with}\quad
\varphi_{0}^2=\frac{\mu^2-m^2}{2\lambda}.
\label{decomp}
\end{equation}
By analyzing the quadratic forms for the two pairs of the fields 
$(\varphi_{1},\varphi_{2})$ and $(\tilde{\varphi}_{1},\tilde{\varphi}_{2})$,
we arrive at the explicit dispersion relations for the charged 
and neutral degrees of freedom,\cite{MS,STV}
\begin{eqnarray}
\omega_{1,2} &=& \sqrt{\mu^{2}+k^{2}}\pm \mu, \\
\tilde{\omega}_{1,2} &=& \sqrt{3\mu^{2}-m^{2} +k^{2}
\pm \sqrt{(3\mu^{2}-m^{2})^{2} + 4\mu^{2}k^{2} }},
\end{eqnarray}
respectively. As is easy to check, here there are only two NG 
bosons with the following dispersion relations, 
\begin{eqnarray}
\omega_{2} &\simeq& \frac{k^{2}}{2\mu}, 
\quad\mbox{for}\quad k\to 0 , \label{NG1} \\
\tilde{\omega}_{2} &\simeq& 
\sqrt{\frac{\mu^{2}-m^{2}}{3\mu^{2}-m^{2}}} k, 
\quad\mbox{for}\quad k\to 0,
\label{NG2}
\end{eqnarray}
which carry the quantum numbers of
$K^{+}$ and $K^{0}$ mesons. The third would-be NG boson, with
the quantum numbers of $K^{-}$, is massive in this model. This
happens despite the fact that the potential part
of Lagrangian density (\ref{L-model}) has three flat directions in the broken
phase, as it should. The splitting between $K^{+}$ 
and $K^{-}$ occurs because of the seesaw mechanism in the kinetic 
part of the Lagrangian density (kinetic
seesaw mechanism).\cite{MS} This mechanism is provided by the
first two terms in expression (\ref{mu-terms}) which, because of
the imaginary unit in front, mix the real and imaginary parts of the
field $\Phi$. Of course this effect is possible only because $C$,
$CP$, and $CPT$ symmetries are explicitly broken in this system
at a nonzero $\mu$. The latter point is reflected in the spectrum of
$K$ mesons even for subcritical values of the chemical potential
$\mu < m$. In that case, there is a splitting of the 
energy gaps (``masses") of
$(K^{-},\bar{K}^{0})$ and $(K^{+},K^{0})$ doublets. While the
gap of the first doublet is equal to $m + \mu$, the gap of the
second one is $m - \mu$. 

Another noticeable point is that while the dispersion relation for 
$K^{0}$ is conventional, with the energy $\omega \sim k$ as the 
momentum $k$ goes to zero, the dispersion relation for 
$K^{+}$ is $\omega \sim k^2$ for small $k$ [see Eqs.~(\ref{NG1})
and (\ref{NG2})]. This fact is in accordance with the Nielsen-Chadha 
counting rule, $N_{G/H} = n_{1} + 2n_{2}$.\cite {NC} Here
$n_{1}$ is the number of NG bosons with the linear dispersion law,
$\omega \sim k$, $n_{2}$ is the number of NG bosons with the quadratic 
dispersion law $\omega \sim k^2$, and $N_{G/H}$ is the number of the 
generators in the coset space $G/H$ (here $G$ is the symmetry group of 
the action and $H$ is the symmetry group of the ground state).

The dispersion relation $\omega \sim k^2$ also implies that, despite
the presence of the condensate, the Landau criterion\cite{superfluid}
fails in this model and the system is not a superfluid. Indeed, recall
that, according to the Landau criterion, superfluidity takes place
for velocities $v < v_c$, where the critical velocity $v_c$ is
the minimum of the ratios $\omega_{i}(k)/k$ taken over all excitation
branches and over all values of momentum $k$. Therefore even the 
presence of a single branch with $\omega \sim k^2$ implies that
$v_c = 0$. 

Does the conventional Anderson-Higgs mechanism survive in the
gauged version of this model despite the absence of one out of 
three NG bosons? This question has motivated the work
where the gauged $\sigma$-model with the chemical potential for
hypercharge was considered.\cite{GMS} The answer to this question 
is positive and we will consider it in the following sections.

\section{Gauged $\sigma$-model}

We will consider the  dynamics in the gauged version of model 
(\ref{L-model}), i.e., the model described by the Lagrangian
density
\begin{eqnarray}
{\cal L}=-\frac{1}{4}F_{\mu\nu}^{(a)}F^{\mu\nu(a)}-\frac{1}{4}B_{\mu\nu}
B^{\mu\nu}+[(D_\mu-i\mu\delta_{\mu0})
\Phi]^\dagger(D^\mu-i\mu\delta^{\mu0})\Phi-V(\Phi^\dagger\Phi),
\label{lagrangian}
\end{eqnarray}
where 
$V(\Phi^\dagger\Phi)= m^2\Phi^\dagger\Phi+\lambda(\Phi^\dagger\Phi)^2$ 
and the covariant derivative $D^\mu=\partial_\mu-igA_\mu- (ig^\prime/2)
B_\mu$. The field $\Phi$ could be taken in the same form as in 
Eq.~(\ref{decomp}) with $\varphi_0$ being the expectation value that 
is determined by minimizing the effective potential. The $SU(2)$ gauge 
fields are given by 
$A_\mu=A_\mu^a \tau^a/2$, where $\tau^a$ are three Pauli matrices, and the
field strength $F_{\mu\nu}^{(a)}=\partial_\mu A_\nu^{(a)}-\partial_\nu 
A_\mu^{(a)}+g\epsilon^{abc}A_\mu^{(b)}A_\nu^{(c)}$. 
$B_\mu$ is the U(1)$_Y$
gauge field with the strength $B_{\mu\nu}=\partial_\mu B_\nu-\partial_\nu
B_\mu$. The hypercharge of the doublet $\Phi$ equals +1.
This model has the same structure as the electroweak theory without
fermions and with the chemical potential for hypercharge $Y$.

We consider two different cases: the case with $g^{\prime} =0$, when
the hypercharge $Y$ is connected with the global $U(1)_{Y}$ symmetry,
and the case with a nonzero $g^{\prime}$, when the $U(1)_{Y}$ symmetry 
is gauged. The main characteristics of these dynamics are the 
following.\cite{GMS} For $m^2 > 0$, the spontaneous breakdown of the
$SU(2) \times U(1)_Y$ symmetry is caused solely by a
supercritical chemical potential $\mu^2 > m^2$. In this case
spontaneous breakdown of the $SU(2) \times U(1)_Y$ is
always accompanied by spontaneous breakdown of both the rotational
symmetry $SO(3)$ [down to $SO(2)$] and the electromagnetic $U(1)_{em}$ 
connected with the electrical charge. Therefore, in this case the
$SU(2) \times U(1)_Y \times SO(3)$ group is broken spontaneously 
down to $SO(2)$. This pattern of spontaneous symmetry breakdown
takes place for both $g^{\prime} =0$ and $g^{\prime} \neq 0$,
although the spectra of excitations in these two cases are different.
Also, the phase transition at the critical point $\mu^2 =m^2$ is
a second order one.

The realization of both the 
NG mechanism and the Anderson-Higgs mechanism
is conventional, despite the unconventional realization 
of the NG mechanism in the original ungauged model (\ref{L-model}).
For $g^{\prime} =0$, there are three NG
bosons with the dispersion
relation $\omega \sim k$,
as should be
in the conventional realization of the breakdown
$SU(2) \times U(1)_Y \times SO(3) \to SO(2)$ when 
$U(1)_Y \times SO(3)$ is a global symmetry. The other
excitations are massive (the Anderson-Higgs mechanism). 
For $g^{\prime} \neq 0$, there are two  
NG bosons with $\omega \sim k$, as should be
when only $SO(3)$ is a global symmetry (the third NG boson
is now ``eaten" by a photon-like combination of fields
$A^{3}_{\mu}$ and $B_{\mu}$ 
that becomes massive). In accordance with the
Anderson-Higgs mechanism, the rest of excitations are massive. 

Since the residual $SO(2)$ symmetry is low, the spectrum
of excitations is very rich. In particular, the dependence of their
energies on the longitudinal momentum $k_3$, directed along the
$SO(2)$ symmetry axis, and on the transverse one, 
${\bf k}_\perp = (k_1,k_2)$, is quite different.
A noticeable point is
that there are two excitation branches, connected with two NG bosons,
that
behave as phonon-like quasiparticles for small momenta
(i.e., their energy $\omega \sim k$) and
as roton-like ones for large momenta $k_3$, i.e., there is a local minimum
in $\omega(k_3)$ for a value of $k_3$ of order $m$
(see the plots of the dispersion relations in 
Secs.~\ref{4} and \ref{6}).
On the other hand, $\omega$ is a monotonically increasing function 
of the transverse momenta. The existence of the roton-like
excitations is caused by the presence of gauge fields [there are
no such excitations in ungauged model (\ref{L-model})]. 
As is well known, excitations with the behavior of such a 
type are present in
superfluid systems and the microscopic origin of these
excitations is a long standing problem.\cite{L,F,superfluid} 
The connection of roton excitations with gauge fields in the gauged
$\sigma$-model is intriguing and could shed light on their
microscopic nature. The presence of these excitations
also suggests that the present model
could be relevant for anisotropic superfluid systems.

In the case of $m^2 < 0$, the spontaneous breakdown of
the $SU(2) \times U(1)_Y$ symmetry takes place even without
chemical potential. Introducing the chemical potential
leads to dynamics similar to that in tumbling gauge
theories.\cite{tumbling} While in tumbling gauge
theories the initial symmetry is breaking down (``tumbling'')
in a few stages with 
increasing the running gauge coupling, in this model two different
stages of symmetry breaking are determined by the values of
chemical potential. When $0 < \mu^2 < \frac{g^2}{16\lambda}|m^2|$,
the $SU(2) \times U(1)_Y$ breaks down to $U(1)_{em}$, and the
rotational $SO(3)$ is exact. In this case, the conventional
Anderson-Higgs mechanism is realized with three gauge bosons being
massive and with no NG bosons.
The presence of $\mu$ leads to splitting of the masses of charged
$\pm 1$ gauge bosons. It is interesting that at this stage the
phenomenon that we call dynamical transmutation of chemical
potential takes place. It is reflected in 
such a rearrangement of the ground state
when $\mu$ ceases to play the role of the chemical potential for 
hypercharge and becomes the chemical potential for
electric charge instead.

The second stage happens when 
$\mu^2$ becomes larger than $\frac{g^2}{16\lambda}|m^2|$. Then
one gets the same breaking pattern as that described above
for $m^2 >0$, with $SU(2) \times U(1)_Y \times SO(3) \to SO(2)$.
The spectrum of excitations is also similar
to that case. 
At last, for all those values of the coupling constants $\lambda$ and $g$
for which the effective potential is bounded from below, the
phase transition at the critical 
point $\mu^2 = \frac{g^2}{16\lambda}|m^2|$ is a second order one.

\section{Model with global $U(1)_Y$ symmetry: $m^2>0$ case}
\label{4}

We begin by making the following general observation.
Let us consider a theory with a chemical potential $\mu$ 
connected with a conserved charge $Q$. Let us introduce
the quantity
\begin{eqnarray}
R_{min} \equiv \mbox{min}(m^2/Q^2),
\label{R}
\end{eqnarray}
where on the right hand side we consider the minimum value 
amongst the ratios
$m^2/Q^2$ for all {\it bosonic} particles with $Q \neq 0$
in this same theory but
{\it without} the chemical potential. Then if $\mu^2 > R_{min}$,
the theory exists only if the spontaneous breakdown 
of the $U(1)_{Q}$ symmetry takes place there. 
Indeed, if the $U(1)_{Q}$ were exact in such a theory,
the partition function, $Z = \rm{Tr}[exp(\mu \hat{Q} - H)/T]$, 
would diverge.\footnote{Similarly as it happens in nonrelativistic 
bose gas with a positive chemical potential connected with the 
number of particles $N$.\cite{LL}} 
Examples of the restriction $\mu^2 < R_{min}$ in relativistic
theories were considered Refs.~\citebk{Kapusta} and \citebk{HW}.

In fact, the value $\mu^2 = R_{min}$ 
is a critical point separating different phases in the theory.
It is important that since in the phase
with $\mu^2 > R_{min}$ the charge $Q$ is not a good quantum
number, $\mu$ ceases to play the role of a chemical
potential determining the density of this charge.
This point was emphasized in Ref.~\citebk{Kapusta}.
There are a few options in this case. If there remains an exact symmetry
connected with a charge $Q^{\prime} = aQ + X$, where 
$a$ is a constant and $X$ represents some other generators,
the chemical
potential will determine the density of the charge $Q^{\prime}$
(we call this phenomena a dynamical transmutation of the chemical 
potential).
Otherwise, it becomes just a
parameter determining the spectrum of excitations and other
thermodynamic
properties of the system (the situation is similar to that taking place
in models when a mass square $m^2$ becomes negative). We will
encounter both these options in model (\ref{lagrangian}).

We begin by considering the case with $g^{\prime} = 0$ 
and $m^2 > 0$. When $\mu^{2} < m^2$, the
$SU(2) \times U(1)_Y \times SO(3)$ symmetry is exact. Of course
in this case a confinement dynamics for three $SU(2)$
vector bosons takes place and it is not under our control. However,
taking $\mu^2 \sim m^2$ and choosing $m$ to be much larger than the
confinement scale $\Lambda_{SU(2)}$, we get controllable
dynamics at large momenta $k$ of order $m$. It includes three
massless vector bosons $A^{a}_{\mu}$ and two doublets,
$(K^{+},K^{0})$ and $(K^{-}, \bar{K}^{0})$. The spectrum of
the doublets is qualitatively the same as that in model
(\ref{L-model}): the chemical potential leads to splitting
the masses (energy gaps) of these doublets and, in tree
approximation, their masses are $m - \mu$ and $m + \mu$,
respectively (see Sec. \ref{ng}). In order to make the tree
approximation to be reliable, one should take $\lambda$
to be small but much larger than the value
of the running coupling $g^4(m)$ related to the scale $m$
[smallness of $g^2(m)$ is guaranteed by the condition
$m \gg \Lambda_{SU(2)}$ assumed above]. The condition
$g^4(m)\ll \lambda \ll 1$ implies that 
the contributions both of vector boson
and scalar loops are small, i.e., there is no Coleman-Weinberg (CW)
mechanism (recall that one should have
$\lambda \sim g^4$ for the CW mechanism).\cite{CW}        

Let us now consider the case with $\mu^2 > m^2 >0$ in detail. 
Since $m^2$ is equal to $R_{min}$ (\ref{R}), there should be 
spontaneous $U(1)_{Y}$ symmetry breaking in this case.   
For $g^{\prime}= 0$,
the equations of motion derived from Lagrangian density  
Eq.(\ref{lagrangian}) read
\begin{eqnarray}
-(D_\mu-i\mu\delta_{\mu0})(D^\mu-i\mu\delta^{\mu0})\Phi-m^2\Phi-2\lambda
(\Phi^\dagger\Phi)\Phi &=& 0,
\label{equation1}\\
\partial^\mu F_{\mu\nu}^{(a)}+g\epsilon^{abc}A^{\mu(b)} F_{\mu\nu}^{(c)}
+ig\left[\Phi^\dagger\frac{\tau^a}{2}\partial_\nu\Phi-\partial_\nu
\Phi^\dagger\frac{\tau^a}{2}\Phi\right] &&\nonumber \\
+\frac{g^2}{2}A_{\nu}^{(a)} \Phi^\dagger\Phi
+2g\mu\delta_{\nu0}\Phi^\dagger\frac{\tau^a}{2}\Phi &=& 0
\label{equation2}
\end{eqnarray}
(since now the field $B_{\mu}$ is free and decouples, we ignore it).
Henceforth we will use the unitary gauge with 
$\Phi^T=(0,\varphi_0+\tilde\varphi_1/\sqrt{2})$.
It is important that the existence of this gauge is based solely on
the presence of $SU(2)$ gauge symmetry, independently of
whether the number of
NG bosons in ungauged model (\ref{L-model}) is conventional or not. 
We will be first
looking for a homogeneous ground state solution (with $\varphi_0$ 
being constant) that does not break the
rotational invariance, i.e., with $A_i^{(3,\pm)}=0$ where
$A_\mu^{(\mp)}=\frac{1}{\sqrt{2}}(A_\mu^{(1)}\pm iA_\mu^{(2)})$.
In this case the equations of motion become
\begin{eqnarray}
\left(i\partial_0A_0^{(+)}+2\mu A_0^{(+)}\right)
\varphi_0=0,&&\\
\left[(\mu-\frac{g}{2}A_0^{(3)})^2-m^2-
2\lambda\varphi_0^2-\frac{ig}{2}
\partial_0A_0^{(3)}+\frac{g^2}{2}A_0^{(+)}A_0^{(-)}\right]
\varphi_0=0,&&\\
g\left(\frac{g}{2}A_0^{(3)}-\mu\right)\varphi_0^2
=0,&&\\
\frac{g^2\varphi_0^2}{2}A_0^{(\pm)}=0.
\end{eqnarray}
Besides the symmetric solution with $\varphi_0=0$,
this system of equations allows the following solution:
\begin{eqnarray}
\varphi_0^2=-\frac{m^2}{2\lambda},\quad
A_0^{(3)}=\frac{2\mu}{g},\quad A_0^{(\pm)}=0.
\label{vacuum2}
\end{eqnarray}
We recall that in the unitary gauge
all auxiliary, gauge dependent, degrees of
freedom are removed. Therefore in this gauge
the ground state expectation values of vector
fields are well defined physical quantities.

Solution (\ref{vacuum2}), describing spontaneous 
$SU(2) \times U(1)_{Y}$ symmetry
breaking, exists only for negative $m^2$. On the other hand, the symmetric
solution with $\varphi_0=0$
cannot be stable in the case of
$\mu^2 > R_{min}= m^2 >0$ we are now interested in. This forces us to
look for a ground state solution that breaks the rotational 
invariance.\footnote{We will get a better insight in the reason why
spontaneous rotational invariance breaking is inevitable for
$\mu^2 > m^2 >0$ from considering the dynamics with
$m^2 < 0$ in Sec.~\ref{5}.} Let us now consider the effective potential 
$V$.
It is obtained from Lagrangian density 
Eq.(\ref{lagrangian}), $V=-{\cal L}$, by setting all field 
derivatives to zero. Then we get
\begin{eqnarray}
V=V_1+V_2,
\label{potential}
\end{eqnarray}
with
\begin{eqnarray}
V_1&=&-\frac{g^2}{2}\left[\left(A_0^{(a)}A_0^{(a)}\right)\left(A_i^{(b)}
A_i^{(b)}\right)-
\left(A_0^{(a)}A_i^{(a)}\right)\left(A_0^{(b)}A_i^{(b)}\right)\right]
\nonumber\\
&+&\frac{g^2}{4}\left(A_i^{(a)}A_i^{(a)}
\right)\left(A_j^{(b)}A_j^{(b)}\right)-\frac{g^2}{4}\left(A_i^{(a)}A_j^{(a)}
\right)\left(A_i^{(b)}A_j^{(b)}\right),\\
V_2&=&(m^2-\mu^2)\Phi^\dagger\Phi+\lambda(\Phi^\dagger\Phi)^2
-2g\mu\Phi^\dagger A_0^{(a)}\frac{\tau^a}{2}\Phi-\frac{g^2}{4}
A_\mu^{(a)}A^{\mu(a)}\Phi^\dagger\Phi.
\end{eqnarray}
We use the ansatz $\Phi^T=(0,\varphi_{0})$ along with
\begin{equation}
A_3^{(+)}=(A_3^{(-)})^*=C\neq0,\quad A_0^{(3)}=D\neq0,
\quad A_{1,2}^{(\pm)}
=A_{0}^{(\pm)}=A_{1,2}^{(3)}=A_{3}^{(3)}=0,
\label{ansatz} 
\end{equation}
that breaks 
spontaneously both rotational symmetry [down to $SO(2)$] and
$SU(2)\times U(1)_Y$ (completely). Substituting
this ansatz into potential (\ref{potential}), we
arrive at the expression 
\begin{equation}
V=-g^2D^2|C|^2-\left(\mu-\frac{gD}{2}\right)^2
\varphi_0^2+\frac{g^2}{2}|C|^2\varphi_0^2+m^2\varphi_0^2+\lambda\varphi_0^4.
\label{potential-with-anzats}
\end{equation}
It leads to the following equations of motion:
\begin{eqnarray}
\left(D^2-\frac{\varphi_0^2}{2}\right)C=0,&&\\
\left(2|C|^2+\frac{\varphi_0^2}{2}\right)D=\frac{\varphi_0^2}{g}
\mu,&&\\
\left[\left(\mu-\frac{gD}{2}\right)^{2} -m^2-2\lambda
\varphi_0^2-\frac{g^2}{2}|C|^2 \right]\varphi_0=0.
\label{anzats-eq}
\end{eqnarray}
One can always take both $g$ and the ground state expectation
value $\varphi_0$ to be positive (recall that we also
take $\mu>0$). Then from the first two equations we obtain
\begin{equation}
 D=\frac{\varphi_0}{\sqrt{2}}>0,\quad 2|C|^2+\frac{\varphi_0^2}{2}=
\frac{\sqrt{2}\mu\varphi_0}{g},
\label{solution-for-A03}
\end{equation}
while the third equation reduces to 
\begin{equation}
\left(\frac{g^2}{4}-2\lambda\right)\varphi_0^2-\frac{3g\mu}{2\sqrt{2}}
\varphi_0+\mu^2-m^2=0.
\label{equation}
\end{equation}
Hence for $\varphi_0$ we get the following solution:
\begin{equation}
\varphi_0=\frac{1}{\sqrt{2}(8\lambda-g^2)}\left[\sqrt{(g^2+64\lambda)
\mu^2+8(8\lambda-g^2)(-m^2)}-3g\mu\right].
\label{solution-varphi}
\end{equation}
It is not difficult to show that for $\mu^2 > m^2 > 0$ both expression 
(\ref{solution-varphi}) for $\varphi_0$ and expression 
(\ref{solution-for-A03}) for $|C|^2$ are positive and, 
for $g^2\leq 8\lambda$, this solution corresponds to the minimum
of the potential. The phase transition at the critical value 
$\mu = m$ is a second order one. 

The situation in the region $g^2> 8\lambda$ is somewhat more
complicated. First of all, in that region the potential 
(\ref{potential-with-anzats}) becomes unbounded from below 
[one can see this after substituting the expression for 
$A_0^{(3)}=D$ from Eq.~(\ref{solution-for-A03}) into the potential]. 
Still, even in that case there is
a local minimum corresponding to solution (\ref{solution-varphi}).
The phase transition is again a second order one. Henceforth
we will consider only the case with $g^2\leq 8\lambda$ when the
potential is bounded from below. 
Notice that for small $g^2 \equiv g^2(m)$ the inequality
$g^2\leq 8\lambda$ is consistent with the condition
$g^4 \ll \lambda$ necessary for the suppression of the
contribution of boson loops, as was discussed above.

\begin{figure}[t]
\centerline{
\psfig{file=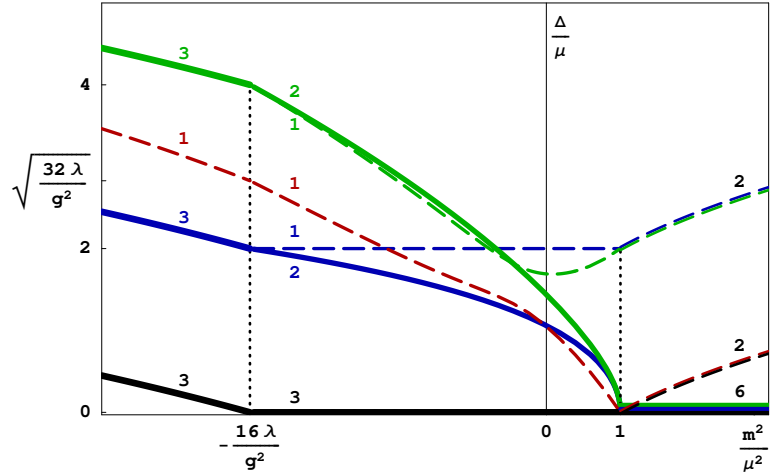,width=4in}}
\vspace*{8pt}
\caption{The energy 
gaps $\Delta$ of excitations in the 
model with global $U(1)_{Y}$ symmetry. The degeneracy factor is specified 
at each gap line.}
\label{fig:spectrum}
\end{figure}

In order to derive the spectrum of excitations, one has to 
introduce small fluctuations $a^{(a)}_{\mu}$ and $\tilde\varphi_1$
about 
the ground state solution in Eq.~(\ref{ansatz})
[i.e., $A^{(a)}_{\mu} = \langle A^{(a)}_{\mu} \rangle + a^{(a)}_{\mu}$
and $\Phi^T=(0,\varphi_0+\tilde\varphi_1/\sqrt{2})]$ and then 
make the expansion in Lagrangian density (\ref{lagrangian})
keeping only quadratic fluctuation terms. 
This analysis of the quadratic form was done
by using {\it MATHEMATICA}.\cite{GMS}
It leads to the energy gap (``mass") spectrum of excitations
shown in Fig.~\ref{fig:spectrum} (both
cases with $m^2 > 0$ and  $m^2 < 0$, considered below in Sec.~\ref{5}, 
are shown there).

Since in the subcritical phase, with $\mu^2 < m^2$, there 
are 10 physical states (6 states connected with three 
massless vector bosons and 4 states connected with the
doublet $\Phi$), there should
be 10 physical states (modes) also in the supercritical
phase. Out of the total 10 modes there exist 3
massless (gapless) NG modes, as should be in the conventional
realization of the spontaneous breakdown of
$SU(2) \times U(1)_Y \times SO(3) \to SO(2)$, when
$U(1)_Y \times SO(3)$ is a global symmetry. The gaps 
$\Delta$ of the excitations are defined as the values of their
energies at zero momentum. They are
\begin{equation}
\begin{array}{ll}
\Delta^2 = 0, & [\times 3] ,\\
\Delta^2 = 2 \mu \phi, & [\times 2] ,\\
\Delta^2 = 2 \mu \phi + 3\phi^2 , & [\times 2] ,\\
\Delta^2 = 4 \mu^2,  & [\times 1] ,\\
\Delta^2 = \delta_{-}^2,& [\times 1] ,\\
\Delta^2 = \delta_{+}^2, & [\times 1] 
\end{array}
\label{gap}
\end{equation}
with the degeneracy factors specified in square brackets. Here
we introduced the following notations:
\begin{eqnarray}
\phi^2 = g^2\varphi_0^2/2\quad \mbox{and} \quad
\delta_{\pm}^2 = F_1\pm\sqrt{F_1^2 - F_2}
\label{delta_pm}
\end{eqnarray}
with
\begin{eqnarray}
F_1 &=& 3\mu^2-m^2 -\frac{7}{2}\mu \phi + 3\phi^2,\\
F_2  &=& 8(3\mu^2-m^2)\phi^2-30\mu \phi^3+9\phi^4.
\end{eqnarray}

The dispersion relations for the NG bosons in the infrared
region are
\begin{eqnarray}
\omega^2 &\simeq& \frac{2\mu-\phi}{2\mu+3\phi}{\bf k}^2+O(k_i^4),\\
\omega^2 &\simeq& \frac{2\mu-\phi}{2\mu+3\phi}\left(
\frac{\phi}{2\mu}{\bf k}_\perp^2+k_3^2\right)+O(k_i^4),\\ 
\omega^2 &\simeq& \frac{ \left(2\mu-\phi\right) \left[
4(\mu^2-m^2)-3\mu\phi\right]}
{\mu \left[ 8(3\mu^2-m^2) -30 \mu\phi+9\phi^2\right]}{\bf k}^2
+O(k_i^4),
\end{eqnarray}
where $\omega \equiv k_0$. The infrared dispersion relations for the other
seven excitations are rather complicated and can be found in the original
paper.\cite{GMS}

\begin{figure}[t]
\centerline{
\psfig{file=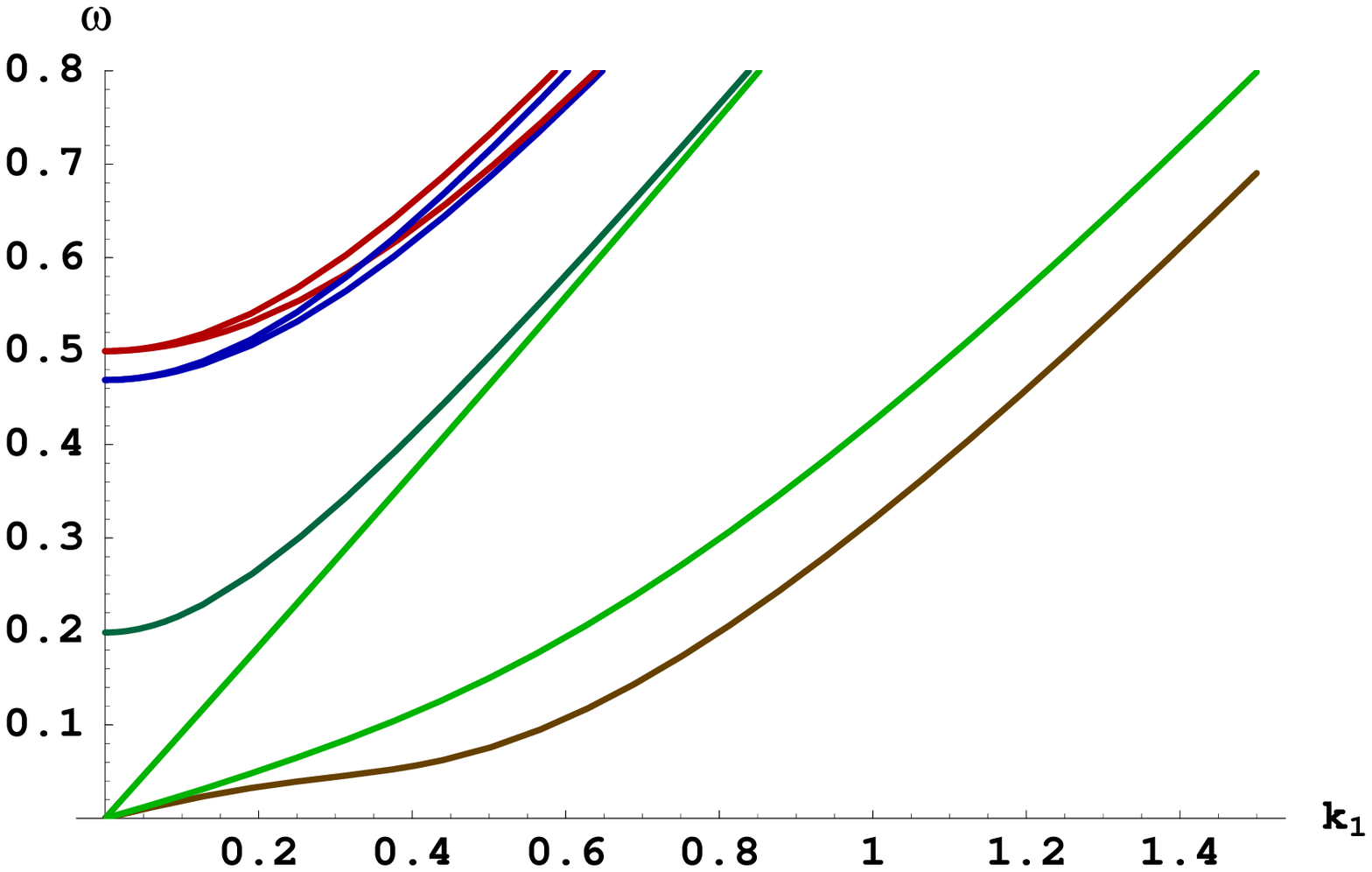,width=2.5in}
\psfig{file=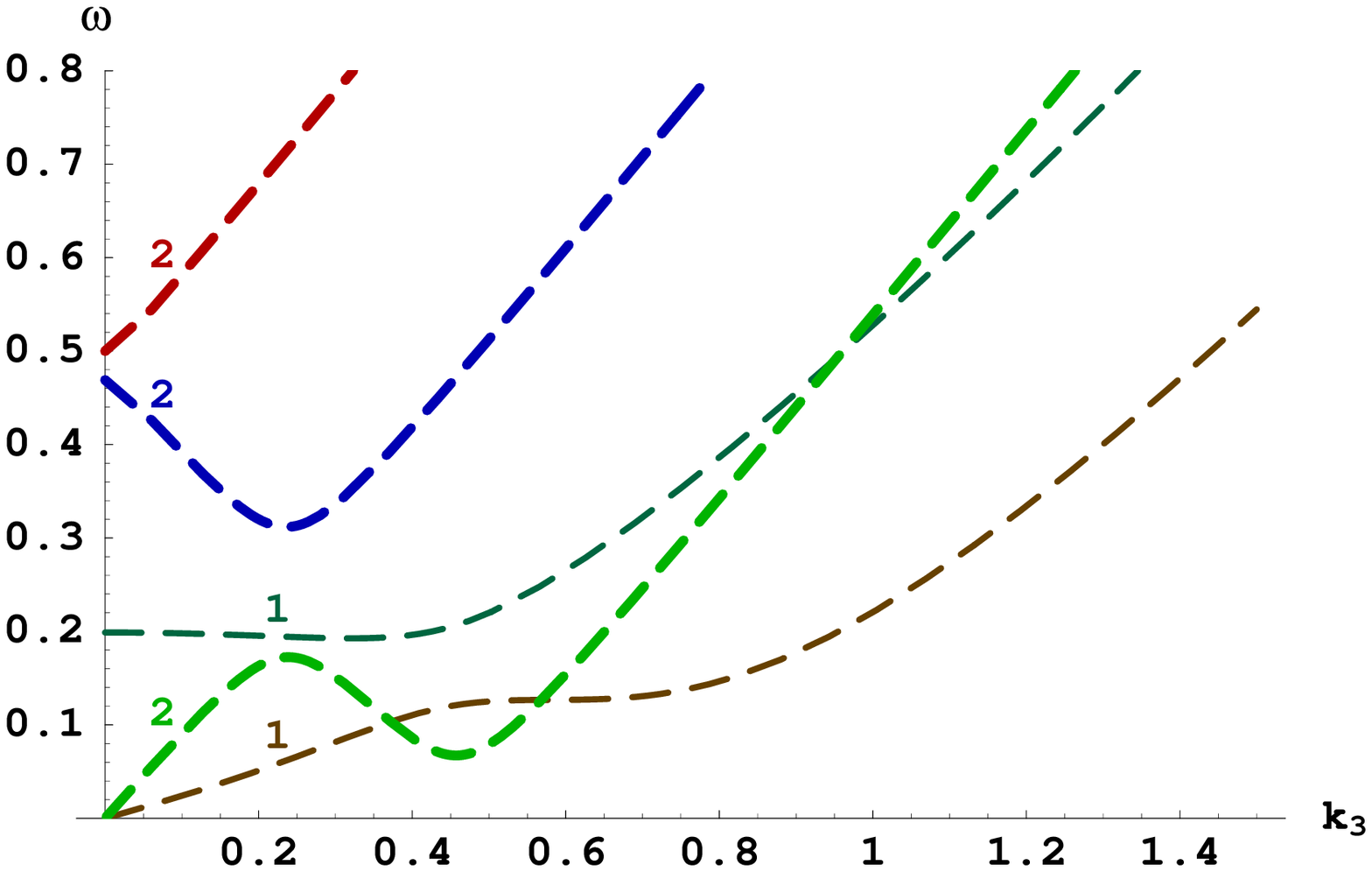,width=2.5in}}
\vspace*{8pt}
\caption{The energy $\omega$ of the 8 light quasiparticle modes as a
function of $k_1$ (solid lines, left panel) and $k_3$ (dashed lines,
right panel). The dispersion relations of two heavy modes are outside
the plot range. The energy and momenta are measured in units of $m$.
The parameters are $\mu/m=1.1$ and $\phi/m=0.1$}
\label{fig-all-0}
\end{figure}

While the analytical dispersion relations in the infrared region 
are quite useful, we performed also numerical calculations to 
extract the corresponding dispersion relations outside the infrared 
region. The results are as follows. In the near-critical region, 
$\mu\to m+0$, the ground state expectation $\phi$ becomes small. 
In this case, one gets 8 light modes, see Eq.~(\ref{gap}). The 
results for their dispersion relations are shown in 
Fig.~\ref{fig-all-0} (the two heavy modes with the gaps of order 
$2\mu$ are not shown there).
The solid and dashed lines represent the energies
of the quasiparticle modes as functions of the transverse
momentum ${\bf k}_\perp = (k_1,0)$ (with $k_3 =0$) 
and the longitudinal momentum $k_3$ (with ${\bf k}_\perp =0$),
respectively. Bold and thin lines correspond 
to double degenerate and nondegenerate modes, respectively. 

There are the following characteristic features of the spectrum.
a) The spectrum with ${\bf k}_\perp =0$ (the right panel in 
Fig.~\ref{fig-all-0}) is much more degenerate 
than that with $k_3 =0$ (the left panel). This
point reflects the fact that the axis of the residual $SO(2)$
symmetry is directed along $k_3$. Therefore the states with
${\bf k}_\perp =0$ and $k_3 \neq 0$ are more symmetric than
those with ${\bf k}_\perp \neq 0$. b) The right panel in 
Fig.~\ref{fig-all-0} contains two NG
branches with local minima at
$k_3 \sim m$, i.e., roton-like excitations. Because there are no such
excitations in ungauged model (\ref{L-model}),
they occur because of the presence of gauge fields.
Since roton-like excitations occur in superfluid systems, 
the present model could be relevant for them. c) The NG 
and Anderson-Higgs mechanisms are conventional in this system. 
In particular, the dispersion relations for three NG bosons
have the form $\omega \sim k$ for low momenta.

When the value of the chemical potential increases, the values of 
masses of all massive quasipatricles become of the same order.
Otherwise, the characteristic features of the dispersion
relations remain the same.

\section{Model with global $U(1)_Y$ symmetry: $m^2<0$ case}
\label{5}

Let us now turn to the case with negative $m^2$. In this case
there is the ground state solution (\ref{vacuum2}) describing
spontaneous breakdown of $SU(2)\times U(1)_{Y}$ down
to $U(1)_{em}$ and preserving the rotational invariance.
In order to describe the spectrum of excitations, 
we make the expansion in Lagrangian
density (\ref{lagrangian}) about this solution.
Introducing as before small fluctuations $a^{(a)}_{\mu}$
[i.e., $A^{(a)}_{\mu} = \langle A^{(a)}_{\mu} \rangle +a^{(a)}_{\mu}$] 
and $\tilde\varphi_1$
[i.e., $\Phi^T=(0,\varphi_0+\tilde\varphi_1/\sqrt{2})]$ 
and keeping only quadratic fluctuation terms, we obtain
\begin{eqnarray}
{\cal L} &\simeq &
\frac{1}{2}f_{0i}^{(a)}f_{0i}^{(a)}- \frac{1}{4}f_{ij}^{(a)}
f_{ij}^{(a)}+2\mu\left(f_{0i}^{(2)}a_i^{(1)}-f_{0i}^{(1)}a_i^{(2)}\right)
+2\mu^2\left(a_i^{(1)}a_i^{(1)}+a_i^{(2)}a_i^{(2)}\right)
\nonumber \\
&+&\frac{g^2\varphi_0^2}{4}a_\mu^{(a)}a^{\mu(a)}
+\frac{1}{2}\partial_\mu\tilde{\varphi}_{1}
\partial^\mu\tilde{\varphi}_{1}
-\frac{1}{2}\left(m^2+6\lambda\varphi_{0}^2\right)\tilde{\varphi}_{1}^2.
\label{L}
\end{eqnarray}
The analysis of the spectrum of eigenvalues of this quadratic
form is straightforward. The dispersion relations for
charged vector bosons are
\begin{eqnarray}
\omega^{2}_{-}  &=& (\sqrt{{\bf k}^2 +\phi^2} + 2\mu)^2,\\
\omega^{2}_{+} &=& (\sqrt{{\bf k}^2 +\phi^2}- 2\mu)^2 ,
\label{charged}
\end{eqnarray}
where $\omega_{+}$ and $\omega_{-}$ are the energies of vector
bosons with $Q_{em} = +1$ and $Q_{em} = -1$, respectively. The
dispersion relations for the neutral vector boson 
and neutral scalars are $\mu$ independent,
\begin{eqnarray}
\omega^{2}_{0} & =& {\bf k}^2 +\phi^2,\\
\omega^{2}_\varphi& =& {\bf k}^2 +4\lambda\varphi_0^2.
\end{eqnarray}
Therefore, the chemical potential leads to splitting the
masses of two charged vector bosons. In fact, it is easy
to check that the terms with the chemical potential in
Lagrangian density (\ref{L}) look exactly as if the
chemical potential $\bar{\mu} = 2\mu$ for the electric
charge $Q_{em}$ was introduced. In other words, 
as a result of spontaneous $U(1)_Y$ symmetry breaking,
the dynamical transmutation of the chemical potential
occurs: the chemical potential for hypercharge transforms
into the chemical potential for electrical charge. Since the
hypercharge of vector bosons equals zero and $\tilde\varphi_1$
scalar is neutral, this transmutation looks quite dramatic:
instead of a nonzero density for scalars, a 
nonzero density for charged vector bosons is generated.
(The factor
$2$ in $\bar{\mu} = 2\mu$ is of course connected with
the factor $1/2$ in $Q_{em} = I^3 + \frac{1}{2}Y$.)

In this phase, the parameter $R_{min}$ defined 
in Eq.~(\ref{R}) equals $\phi^2 =\frac{g^2}{4\lambda}|m^2|$,
i.e., it coincides with the square of the mass of vector
bosons in the theory without chemical potential.
Therefore, as the chemical potential $\bar{\mu}^2$ becomes larger
than $\phi^2 =R_{min}$, a new phase transition should happen.
And since for $\bar{\mu}^2 = \phi^2$ vector bosons
with charge $+1$ become gapless [see Eq.~(\ref{charged})],
one should expect that
this phase transition is 
triggered by generating a condensate of
charged vector bosons. 

And such a condensate arises indeed. It is not
difficult to check that when 
$\bar{\mu}^2 > \bar{\mu}^{2}_{cr} \equiv  \frac{g^2}{4\lambda}|m^2|$ ,
the ground state solution with ansatz (\ref{ansatz})
occurs. The parameters $C$, $D$, and $\varphi_0$ are
determined from Eqs.~(\ref{solution-for-A03}) and
(\ref{solution-varphi}), respectively. For 
$\bar{\mu}^2 >\bar{\mu}^{2}_{cr}$,
both expression
(\ref{solution-varphi}) for $\varphi_0$ and
expression (\ref{solution-for-A03}) for
$|C|^2$ are positive and,
for $g^2\leq 8\lambda$, this solution corresponds to the 
global minimum
of the potential. The phase transition at the critical value
$\bar{\mu}^2 = \bar{\mu}^{2}_{cr}$
is a second order one.\footnote{As was shown
above, for 
$g^2 > 8\lambda$,
the potential (\ref{potential-with-anzats}) is unbounded from
below, and we will not consider this case.}
The spectrum of excitations in the supercritical phase with
$\bar{\mu}^2 > \bar{\mu}^{2}_{cr}$
is similar to the spectrum in the
case of positive $m^2$ and $\mu^2 > m^2$ (see 
Fig.~\ref{fig:spectrum}).

Therefore, for $m^2 <0$ the breakdown of the initial symmetry 
is realized
in two steps, similarly as it takes place in tumbling gauge
theories.\cite{tumbling}
At the first stage, the initial symmetry $SU(2) \times U(1)_{Y}
\times SO(3)$ breaks down to $U(1)_{em} \times SO(3)$, and then,
at the second stage, the symmetry $U(1)_{em} \times SO(3)$
breaks further to $SO(2)$.
Now we can understand more clearly 
why in the case of positive $m^2$ considered above
the breakdown of the initial symmetry is realized in
one stage. The point is that in that case 
vector bosons in the theory without chemical potential are
massless. Therefore, while for the
chemical potential  connected with hypercharge $R_{min} =m^2$
is positive, $R_{min} =0$
for the chemical
potential connected with electrical charge $Q_{em}$ there. 
This in turn implies that in that case there is no way for 
increasing 
$R_{min}$ through the process of
the transmutation of the chemical potential as it happens
in the case of negative $m^2$. Therefore for $m^2 >0$
the phase in which both the
$U(1)_{em}$ symmetry and the rotational symmetry
are broken occurs at once as $\mu^2$ becomes larger
than $m^2$. 

\section{Model with gauged $U(1)_Y$ symmetry}
\label{6}

Let us now describe the case with $g^{\prime} \neq 0$.
In this case the $U(1)_{Y}$ symmetry is local and one should
introduce a source term $B_{0}J_{0}$ in Lagrangian density 
(\ref{lagrangian}) in order to make the system neutral
with respect to hypercharge $Y$. This is necessary since
otherwise in the system with a nonzero chemical potential
$\mu$ the thermodynamic equilibrium could not be established.
The value of the background hypercharge density $J_{0}$
(representing very heavy particles) is determined from
the condition $\langle{B_0}\rangle =0$.\cite{Kapusta}

After that, the analysis follows closely to that of the case with 
$g^{\prime} =0$. Because of the additional vector boson $B_{\mu}$,
there are now 12 quasiparticles in the spectrum. The sample of 
spontaneous $SU(2) \times U(1)_Y \times SO(3)$ symmetry breaking
is the same as for $g^{\prime} =0$ both for $m^2 \geq 0$
and $m^2 < 0$, with a tumbling-like scenario for the latter.
However, for supercritical values of the chemical potential, there 
are now only two gapless NG modes (the third one is ``eaten" 
by a photon-like combination of fields $A^{3}_{\mu}$ and $B_{\mu}$ 
that becomes massive). Their dispersion relations in infrared read
\begin{eqnarray}
\omega^2 &\simeq& \frac{2\mu-\phi}{2\mu+3\phi}{\bf k}^2+O(k_i^4),\\
\omega^2 &\simeq& \frac{2\mu-\phi}{2\mu+3\phi}\left(
\frac{\phi}{2\mu}{\bf k}_\perp^2+k_3^2\right)+O(k_i^4).
\end{eqnarray}
The rest 10 quasiparticles are gapped. The masses (gaps) of the
massive states are
\begin{eqnarray}
\Delta^2 = 2 \mu \phi + 3\phi^2 , &\quad & [\times 2] ,\\
\Delta^2 = \mu \phi + \frac{\phi_b^2 }{2}
-\sqrt{\left(\mu \phi -\frac{\phi_b^2}{2}\right)^2+\phi^2 \phi_b^2}, 
&\quad & [\times 2] ,
\label{goes_to_0} \\
\Delta^2 = \mu \phi + \frac{\phi_b^2 }{2}
+\sqrt{\left(\mu \phi -\frac{\phi_b^2}{2}\right)^2+\phi^2 \phi_b^2}, 
&\quad & [\times 2] ,\\
\Delta^2 = 2 \mu^2+ \frac{\phi_b^2 }{2}
-\sqrt{\left( 2 \mu^2 -\frac{\phi_b^2}{2}\right)^2+2\mu\phi\phi_b^2},  
&\quad & [\times 1] ,\\
\Delta^2 = 2 \mu^2+ \frac{\phi_b^2 }{2}
+\sqrt{\left( 2 \mu^2 -\frac{\phi_b^2}{2}\right)^2+2\mu\phi\phi_b^2},  
&\quad & [\times 1] ,\\
\Delta^2 = \delta_{-}^2,&\quad & [\times 1] ,\\
\Delta^2 = \delta_{+}^2, &\quad & [\times 1],
\label{gap1}
\end{eqnarray}
where $\phi_b^2 =(g^{\prime})^2\varphi_0^2/2$
with $\varphi_0^2$ given in Eq.~(\ref{solution-varphi}), and 
$\delta_{\pm}$ were introduced in Eq.~(\ref{delta_pm}). Notice
that the double degenerate gap in Eq.~(\ref{goes_to_0}) goes to 
zero together with $g^{\prime}$, i.e., these two
degrees of freedom describe the two transverse states
of massless vector boson $B_{\mu}$ in this limit. 

\begin{figure}[t]
\centerline{
\psfig{file=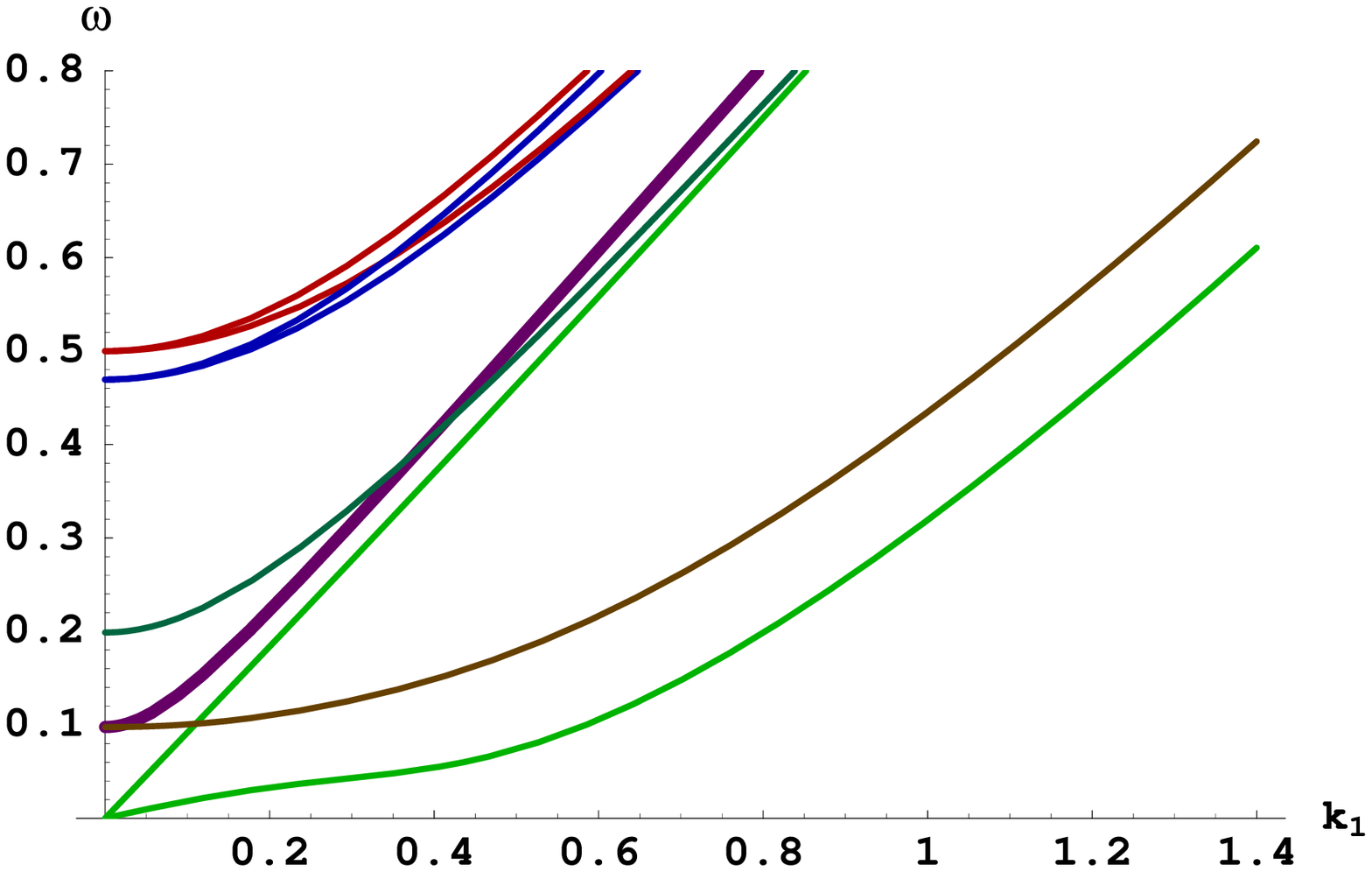,width=2.5in}
\psfig{file=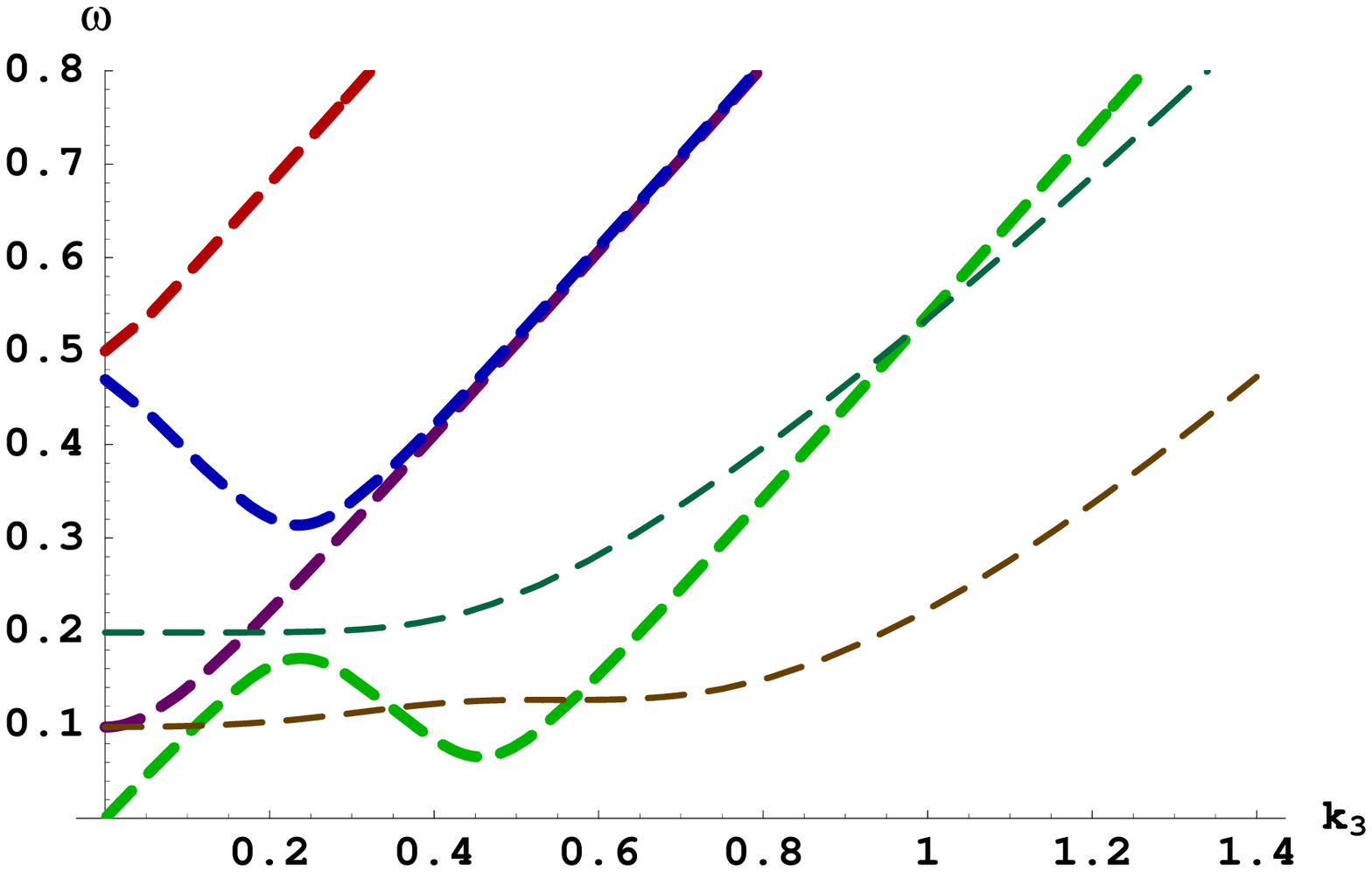,width=2.5in}}
\vspace*{8pt}
\caption{The energy $\omega$ of the 10 light quasiparticle modes as a 
function of $k_1$ (solid lines, left column) and $k_3$ (dashed lines,
right column). The dispersion relations of two heavy modes are outside
the plot ranges. The energy and momenta are measured in units of $m$.
The parameters are $\mu/m=1.1$, $\phi/m=0.1$ and
$\phi_b/m=0.1$.}
\label{fig-all}
\end{figure}

The dispersion 
relations for 10 massive particles
are quite complicated. Therefore in order
to extract the corresponding dispersion relations, numerical
calculations were performed.\cite{GMS} 
These dispersion relations are shown 
in Fig.~\ref{fig-all}. 
Bold and thin lines correspond 
to double degenerate and nondegenerate modes, respectively.
As one can see, the two branches connected with gapless NG
modes contain roton-like excitations at $k_3 \sim m$. Other
characteristic features of the spectrum are also similar
to those of the spectrum for the case with $g^{\prime}=0$
shown in Fig.~\ref{fig-all-0}.
\section{Summary}

The dynamics of the linear $\sigma$-model with a finite chemical
potential for hypercharge is new, rich and, that is very important,
controllable. In the ungauged version of the model, a dynamics with an
abnormal number of NG bosons realizes and, despite the
presence of a condensate, there is no superfluidity in the system.
The richness of the spectrum of excitations in the gauged version of the 
model 
is provided by the coexistence of 
spontaneous breakdown of rotational
symmetry and that
of the electromagnetic $U(1)_{em}$. It is noticeable that 
the spectrum contains   
roton-like excitations. Their connection with
gauge fields is intriguing and deserves further study.

The coexistence of
spontaneous breakdown of the rotational symmetry with that
of the electromagnetic
$U(1)_{em}$ is provided by a
condensate of vector charged bosons.
The possibility of a condensation of vector bosons was
considered in the literature in models different from the present
one.\cite{linde,shabad,lenaghan,sannino} The advantage of the
linear $\sigma$-model without fermions is in its simplicity.
The model admits a controllable dynamics that puts
the creation of a vector condensate on a solid ground and
allows to study the spectrum of excitations in detail.

One can expect that the phenomena discussed here should exist 
in a wide class of relativistic models at finite density. 
In connection with that, it is noticeable that recently
a general approach to the description of systems with an
abnormal number of NG bosons has been developed by 
Nambu.\cite{N} Also, in a very recent paper\cite{BE} the phenomenon with 
an abnormal number of NG bosons was revealed
in the extended Nambu-Jona-Lasinio model with spontaneous
breakdown of the color $SU(3)_{c}$ down to $SU(2)_{c}$.
This model can be relevant for color superconductivity in
dense QCD with two light quark flavors. 

The linear $\sigma$-model 
continues to teach and surprise us.

\section*{Acknowledgments}
V.P.G. and V.A.M. are grateful for support from the Natural 
Sciences and Engineering Research Council of Canada.
The work of I.A.S. was supported by Gesellschaft f\"{u}r 
Schwerionenforschung (GSI) and by Bundesministerium f\"{u}r 
Bildung und Forschung (BMBF).

\section*{Note Added}
After this brief review was published in MPLA, an interesting
paper \cite{CRZ} has appeared in which
roton-like excitations are studied in non-commutative dynamics.

\end{document}